\documentstyle{article}
\newcommand{\be}{\begin{equation}}
\newcommand{\ee}{\end{equation}}
\newcommand{\bea}{\begin{eqnarray}}
\newcommand{\eea}{\end{eqnarray}}
\newcommand{\ba}{\begin{array}}
\newcommand{\ea}{\end{array}}

\def\bbox{{\,\lower0.9pt\vbox{\hrule \hbox{\vrule height 0.2 cm
\hskip 0.2 cm \vrule height 0.2 cm}\hrule}\,}}
\newcommand{\dsl}{\pa \kern-0.5em /}

\newcommand{\nn}{\nonumber \\}

\def\bfo{\mbox{\boldmath $\omega$}}
\def\bfO{\mbox{\boldmath $\Omega$}}
\def\bfn{\mbox{\boldmath $\nabla$}}
\def\bfs{\mbox{\boldmath $\sigma$}}

\def\today{\ifcase\month\or
  January\or February\or March\or April\or May\or June\or
  July\or August\or September\or October\or November\or December\fi
 \space\number\day, \number\year}

\input amssym.def
\input amssym.tex
\font\mybb=msbm10 at 10pt
\def\bb#1{\hbox{\mybb#1}}

\def\bE {\bb{E}}

\def\bC {\bb{C}}

\def\bfo{\mbox{\boldmath $\omega$}}
\def\bfO{\mbox{\boldmath $\Omega$}}
\def\bfn{\mbox{\boldmath $\nabla$}}

\begin{document}


\begin{titlepage}
\vfill
\begin{flushright}
QMW-PH-00-04\\
KCL-TH-00-36\\
DAMTP-2000-69\\
hep-th/0007124\\
\end{flushright}


\vfill
\begin{center}
\baselineskip=16pt
{\Large\bf Supersymmetric Intersecting Domain Walls\\
 in Massive Hyper-K\"ahler Sigma Models}
\vskip 0.3cm
{\large {\sl }}
\vskip 10.mm
{\bf ~Jerome P. Gauntlett$^{*,1}$, ~David Tong$^{\sharp,2}$\\
 and  ~Paul K. Townsend$^{\dagger,3}$ } \\
\vskip 1cm
{\small
$^*$
  Department of Physics\\
  Queen Mary and Westfield College,\\
  Mile End Rd, London E1 4NS, UK\\
}
\vspace{6pt}
{\small
 $^\sharp$
Dept of Mathematics\\
Kings College, The Strand \\
London, WC2R 2LS, UK\\
}
\vspace{6pt}
{\small
 $^\dagger$
DAMTP,\\
Centre for Mathematical Sciences, \\
Wilberforce Road, \\
Cambridge CB3 0WA, UK\\
}
\end{center}
\vfill
\par

\begin{center}
{\bf ABSTRACT}
\end{center}
\begin{quote}

The general scalar potential of D-dimensional massive sigma-models with eight
supersymmetries is found for $D=3,4$. These sigma models typically admit 1/2
supersymmetric domain wall solutions and we find, for a particular 
hyper-K\"ahler target, exact 1/4 supersymmetric static solutions 
representing a non-trivial intersection of two domain walls. We also
show that the intersecting domain walls can carry Noether
charge while preserving 1/4 supersymmetry. We briefly discuss an 
application to the D1-D5 brane system. 
\vfill
\hrule width 5.cm
\vskip 2.mm
{\small
\noindent $^1$ E-mail: j.p.gauntlett@qmw.ac.uk \\
\noindent $^2$ E-mail: tong@mth.kcl.ac.uk \\
\noindent $^3$ E-mail: p.k.townsend@damtp.cam.ac.uk \\
}
\end{quote}
\end{titlepage}
\setcounter{equation}{0}
\section{Introduction}

Supersymmetric four-dimensional (D=4) field theories with multiple isolated
supersymmetric  vacua typically admit 1/2 supersymmetric (BPS) domain wall
solutions that interpolate between these vacua. The Wess-Zumino 
model provides a
simple example with  N=1 supersymmetry; the critical points of its
superpotential are isolated supersymmetric vacua and the minimal tension domain
walls that separate them are BPS \cite{AT1,CQR}. Examples with N=2 
supersymmetry are
provided by certain massive sigma-models; these have no superpotential but they
do have a potential proportional to the square of a tri-holomorphic Killing
vector field (KVF) of the hyper-K\"ahler (HK) target space \cite{AGF}. 
Fixed points of this KVF are supersymmetric vacua and models 
with multiple fixed
points admit 1/2 supersymmetric domain walls \cite{AT2}. 
Recently it has been discovered that N=1 models for which the 
superpotential has
at least three critical points typically admit {\sl intersecting} domain wall
solutions preserving 1/4 supersymmetry \cite{GT,CHT,saffin,OINS,BtV}. The aim of this paper is to
show that there is a class of massive N=2 supersymmetric sigma-models that
similarly admit 1/4 supersymmetric intersecting domain wall solutions.

To see why such configurations might be expected, 
consider a massive sigma model
with a  target space that is the direct product of two 4-dimensional HK target
spaces, each admitting a tri-holomorphic KVF.  We then have two non-interacting
sigma models, each with its potential proportional to a tri-holomorphic KVF,
and a domain wall solution of one model can be superposed with a
domain wall solution of the other model. Since each domain wall separately
preserves 1/2 supersymmetry, experience suggests that some superposition will
preserve 1/4 supersymmetry, and we shall confirm this intuition. Of course,
this example is a trivial one because the intersection is purely geometrical.
We are principally interested in massive sigma models with {\sl irreducible}
target spaces for which simple superposition fails; in this case any
geometrical intersection must also be a physical one in the sense that 
each domain wall must deform the other near the point of intersection.  
Nevertheless, the trivial case suggests that we should
consider HK target spaces admitting at least two linearly independent
triholomorphic KVFs, which requires the HK target space to be at least
8-dimensional. 

We shall consider the class of toric HK $4n$-metrics, which admit $n$
linearly independent commuting KVFs. In coordinates $(\psi_I,{\bf X}^I)$
($I=1,\dots,n$), these metrics are determined by an $n\times n$ symmetric matrix
$U$ with entries $U_{IJ}$ that are functions only of the $3n$ coordinates ${\bf
X}^I$. The metric takes the form 
\be\label{metric}
ds^2 = U_{IJ}d{\bf X}^I \cdot d{\bf X}^J + 
U^{IJ}(d\psi_I + A_I)(d\psi_J + A_J)\, ,
\ee
where $U^{IJ}$ are the entries of $U^{-1}$ and
\be
A_I= d{\bf X}^J\cdot \bfo_{JI}\, ,\qquad 
\bfn_{(J} \times \bfo_{K)I} = \bfn_J U_{KI}\, .
\ee
This last equation implies that $\bfn_{[J} U_{K]I} =0$. The triplet of Kahler
2-forms is
\be
\bfO = (d\psi_I + A_I)d{\bf X}^I -{1\over2}U_{IJ}d{\bf X}^I \times d{\bf X}^J
\ee
where the wedge product of forms is implicit. It follows that the $n$ Killing
vector fields 
\be\label{trihol}
k^I = \partial/\partial\psi_I
\ee
are tri-holomorphic. 

For $n=1$ we get the well-known multi-centre 4-metrics; the simplest two-centre
case is the Eguchi-Hanson 4-metric, and in this case the kink solution of the
corresponding massive sigma model is known explicitly \cite{AT2}. Consider now
the $n=2$ HK 8-metrics. If the $2\times 2$ matrix $U$ is diagonal then
the 8-metric is a direct product of two HK 4-metrics; otherwise, it is
irreducible. In either case, a  potential equal to the square of a
tri-holomorphic KVF must take the form
\be\label{potone}
V = {1\over2}|\lambda_I k^I|^2 
\ee
for some constants $\lambda_I$. When $U$ is diagonal the KVFs $k^1$ and $k^2$
are orthogonal, so the potential can be written as 
\be\label{potspecial}
V= {1\over2}\lambda_1^2 |k^1|^2 + {1\over2}\lambda_2^2|k^2|^2\, ,
\ee
as expected for a Lagrangian that is the sum of two non-interacting massive
sigma-models with 4-dimensional HK target spaces. For this special case, the
potential (\ref{potone}) is the most general one compatible with the N=2
D=4 supersymmetry. As discussed above, it leads to a model that allows a
trivial, but 1/4 supersymmetric, intersection of kink domain walls. One might
hope that a sigma-model with an {\sl irreducible} HK 8-metric and a
potential of the form (\ref{potone}) might admit a supersymmetric 
solution representing a
genuinely physical intersection of domain walls, but this seems not to be the
case.

Fortunately, when $U$ is non-diagonal the potential (\ref{potone}) is {\sl not}
the most general one allowed by N=2 D=4 supersymmetry. As we shall show, the 
general potential is the the sum of squares of {\sl two} independent
tri-holomorphic KVFs. For the toric HK models this means that
\be
V = {1\over2}|\lambda_I k^I|^2 + {1\over2}|\rho_I k^I|^2
\ee
where $\lambda_I$ and $\rho_I$ are two constant $n$-vectors. Equivalently,
\be\label{pottwo}
V = {1\over2}\mu^2_{IJ} U^{IJ}
\ee
where
\be
\mu^2_{IJ}= \lambda_I\lambda_J + \rho_I\rho_J\, .
\ee
In other words, the constants $\mu^2_{IJ}$ are the entries of a constant
symmetric mass-squared matrix $\mu^2$ of maximal rank $2$. 
For $n=2$, this means
that $\mu^2$ is an arbitrary constant symmetric matrix, so the general scalar
potential of a sigma model with a generic 8-dimensional HK target space depends
on {\sl three} parameters. For the special case of a product HK 8-manifold the
number of parameters is reduced to two; these can be taken to be the constants
$\lambda_1$ and $\lambda_2$ in (\ref{potspecial}), which is therefore the
general potential. In contrast, for an irreducible 8-manifold the potential
(\ref{potone}) is the special case of (\ref{pottwo}) for which the
matrix $\mu^2$ has non-maximal rank. 

For any massive sigma model with an 8-dimensional HK target space the
supersymmetric vacua for {\sl generic} mass-squared matrix $\mu^2$
are in 1-1 correspondence with a finite set of points in $\bE^3\times \bE^3$ 
at which the $2\times 2$ matrix $U^{-1}$ equals the zero matrix\footnote{This
is true both classically and in the quantum theory because instanton solutions
interpolating between these vacua are Euclidean kinks with an even number of
fermion zero modes.}. These vacua will be interconnected via 1/2 supersymmetric
domains. As we shall see, the additional parameter in the potential for
irreducible target spaces allows the construction of models admitting 1/4
supersymmetric domain wall intersections.  We shall discuss a particular
one-parameter family of such models in detail, obtaining an explicit and
{\sl exact} 1/4 supersymmetric solution representing a {\sl non-trivial}
intersection of two domain walls.  

The above result for the general potential compatible with 
D=4 N=2 supersymmetry
is a special case of an analogous result for the general scalar potential of a
D-dimensional sigma model with eight supersymmetries. This potential is the sum
of squares of $(6-D)$ linearly independent commuting tri-holomorphic KVFs. This
was conjectured by Bak et al. \cite{Bak} on the basis of their  result that the
D=1 sigma model, i.e. quantum mechanics, admits a potential equal to the sum of
squares of five independent commuting tri-holomorphic KVFs, and the observation
that this might be explained by a five-fold non-trivial 
dimensional reduction on
$T^5$. Indeed, a non-trivial dimensional reduction 
of the, necessarily massless,
D=6 supersymmetric sigma model is known to yield a D=5 supersymmetric sigma
model with a superpotential equal to the square of a tri-holomorphic KVF
\cite{ST}, but it was not previously appreciated that a further non-trivial
dimensional reduction using other tri-holomorphic KVFs might yield more general
potentials. Here we establish, by a direct determination of the conditions
implied by N=2 D=4 and N=4 D=3 supersymmetry that this procedure indeed yields
the general scalar  potential compatible with eight supersymmetries.

Of course, a potential can be constructed from $(6-D)$ independent
commuting tri-holomorphic KVFs only in those models for which the space spanned
by a maximally commuting set of tri-holomorphic KVFs 
has a dimension of at least
$(6-D)$. For the $4n$-dimensional toric HK manifolds this means that we need
$n\ge (6-D)$, but it is reasonable to suppose that most of the physics is
captured by the minimal $n=6-D$ case. This means that we should consider $n=2$
in D=4, as we have been doing, but that we should consider $n=3$ in D=3. One
might suppose that there will now be triple intersections of domain walls
preserving only 1/8 supersymmetry, but the reduced space dimensionality does
not allow it. Instead, one can use the additional freedom to find {\sl charged}
1/4-supersymmetric intersecting domain wall solutions. These have a D=4
interpretation as intersecting walls with a wave along the string
intersection; the momentum along the string becomes the charge on reduction to
D=3. These charged D=3 solutions are stationary rather than static; i.e. they
are time-dependent but such that the energy density is time independent. 
They include, as a special case, new stationary charged domain wall solutions
preserving 1/4 supersymmetry (in contrast to the Q-kinks which preserve 1/2
supersymmetry). In D=3 one can thus break 1/2 supersymmetry to 1/4
supersymmetry {\sl either} by intersection  with another domain wall {\sl
or} by the addition of charge, or both. 

We begin with a derivation of the scalar potentials allowed in supersymmetric
HK sigma models. We then focus on the toric HK manifolds and derive the
BPS equations satisfied by sigma model configurations that preserve some
fraction of supersymmetry. We then apply these results to find both
static and stationary intersecting brane configurations. In the concluding
section we briefly discuss the relevance of these results to the  D1-D5
brane system in type IIB string theory.

\section{Hyper-K\"ahler supersymmetric sigma models}

Massless supersymmetric hyper-Kahler sigma models exist in all spacetime
dimensions $D\le6$. All such models in $D\ge3$ have  
eight supercharges. This is
obvious for $D=5,6$ because the minimal spinor 
in these dimensions has eight real
components. In D=3 the minimal spinor has 2 components, but each of the three
complex structures yields an additional supersymmetry, leading to an N=4 D=3
supersymmetric sigma model, which again has eight supercharges. In D=4 we have
an N=2 model with two 4-component spinor supercharges. 

A D=6 supersymmetric sigma model is necessarily massless as the chirality of
the hypermultiplet spinor allows neither a mass term, nor a Yukawa coupling to
the scalar fields. This argument does not apply in D=5 however, and a
`massive' D=5 supersymmetric sigma model can be found by non-trivial
dimensional reduction\footnote{This type of 
dimensional reduction was introduced
by  Scherk and Schwarz \cite{SS} but their idea was to use it to obtain {\sl
supersymmetry breaking} mass terms, whereas we are using it here to obtain 
supersymmetry preserving mass terms, so we avoid the use of the
terminology `Scherk-Schwarz' reduction.} of the D=6 model if the 
HK target space
admits a tri-holomorphic KVF $\zeta$. Specifically, given
$4n$ real fields $\phi^X$, and a mass parameter $\mu$, one sets
\be
{\partial\phi^X\over\partial x^5} = \mu\zeta^X \qquad (X=1,\dots,4n) 
\ee
where $x^5$ is the coordinate of the 5th spatial direction on which 
we reduce\footnote{The corresponding $x^5$-dependence of the fermion fields
can be found by interpreting this equation as a superfield equation.}. This produces a scalar potential proportional to the square of
the norm of $\zeta$. Given a further tri-holomorphic KVF we could repeat this
procedure (as suggested in \cite{Bak}) to obtain a potential in D=4 that is the
sum of squares of two tri-holomorphic KVFs, and then in D=3 to obtain a
potential that is the sum of three tri-holomorphic KVFs, but one should expect
supersymmetry to require some restriction on the sets of tri-holomorphic KVFs
that can be used for this purpose. Here we shall show, by a direct construction
of the general massive D=3 and D=4 supersymmetric sigma models with eight
supercharges, that the tri-holomorphic KVFs used in this procedure must commute.
The same calculation also shows that this procedure yields the general scalar
potential. 

We shall first review some features of HK manifolds that will be needed in the
construction, and the corresponding D=6 massless sigma models. We then turn to
the construction of the D=4 massive HK sigma models, followed by the D=3
case. 

\subsection{Hyper-Kahler manifolds}

As above, we let $\phi^X$ be the real coordinates of a $4n$-dimensional HK
manifold. The HK metric can be written in terms of a vielbein $f_X^{ia}$
transforming in the $(2,2n)$ representation of $Sp_1\times Sp_n$, as
\be
g_{XY} = f_X^{ia}f_Y^{jb}\Omega_{ij}\varepsilon_{ab}  \qquad (i=1,\dots,2n;\
a=1,2)
\ee
where $\Omega_{ij}$ is the real antisymmetric $Sp_n$-invariant tensor
(numerically equal to $\Omega^{ij}$) and $\varepsilon_{ab}$ is the real
antisymmetric invariant tensor of $Sp_1\cong SU(2)$. The vielbein obeys the
pseudo-reality condition
\be
(f_X^{ia})^* = f_{Xia} \equiv f_X^{jb}\Omega_{ji}\varepsilon_{ba}\, .
\ee
We may define
\be
f^X_{ia} \equiv g^{XY}f_{Yia}
\ee
where $g^{XY}$ is the inverse of the HK metric, in which case 
\be\label{inverse}
f_X^{ia}f^X_{jb} \equiv \delta^i_j \delta^a_b\, .
\ee
Note the further identity
\be\label{fiden}
f_X^{ia} f^Y_{ib} = {1\over2}\left[ \delta_X^Y\delta_b^a + i \bfs_b{}^a \cdot
{\bf I}_X{}^Y\right]
\ee
where $\bfs$ is the triplet of Hermitian Pauli matrices and ${\bf I}$ is the
triplet of complex structures. Given (\ref{inverse}) this identity can be
established by showing that
\be
{\bf I}_X{}^Y = -if_X^{ia}f^Y_{ib} \bfs_a{}^b\, .
\ee
A simple computation using this formula shows that
\be
[{\bf n}\cdot {\bf I}][{\bf m}\cdot {\bf I}] = - {\bf n}\cdot {\bf m} + {\bf
n}\times{\bf m}\cdot {\bf I}
\ee
Taking ${\bf n}$ and ${\bf m}$ to be any two of three orthonormal vectors one
recovers the algebra of quaternions. 

Note that the triplet of K\"ahler 2-forms associated with this quaternionic
structure is
\be
\bfO= -{i\over2}d\phi^X\wedge d\phi^Y\, f_{X}^{ia}f_{Yib} \bfs_a{}^b\, .
\ee
A tri-holomorphic KVF $\zeta$ is one for which
\be
{\cal L}_\zeta \bfO =0\, .
\ee
Given the Killing condition $\zeta_{(X;Y)}=0$, this is equivalent to the
constraint
\be
\zeta_{[ia;j]b}=0
\ee
where $\zeta^{ia} = \zeta^Xf_X^{ia}$. Together with the Killing condition, this
implies that the only non-vanishing component of $\zeta_{X;Y}$ is the symmetric
second-rank $Sp_n$ tensor $\zeta_{ia;j}{}^a$. 

Finally, we observe that the spin-connection, defined by requiring 
\be
{\cal D}_{[X}f_{Y]}{}^{ia} =0
\ee
takes values in $Sp_n$ for a HK manifold, and therefore has
the form
\be
\omega_X{}^{ia}{}_{jb} = \omega_X{}^i{}_j \, \delta^a_b 
\ee
with $\omega_X{}^i{}_i =0$. It follows that the curvature tensor 
takes the form
\be
R_{ia\, jb\, kc\, ld} = R_{ijkl}\epsilon_{ab}\epsilon_{cd}\, ,
\ee
where $R_{ijkl}$ is a totally-symmetric $Sp_n$ tensor.

\subsection{D=6 HK sigma models}

The $D=6$ sigma model has a Lagrangian given, in $SU^*(4)$ spinor
notation, by \cite{ST}
\be
{\cal L}= -{1\over8}g_{XY}\partial^{\alpha\beta} 
\phi^X \partial_{\alpha\beta} \phi^Y -i\chi_{\alpha i}{\cal
D}^{\alpha\beta}
\chi_\beta^i - {1\over 12}
R_{ijkl}\chi_\alpha^i\chi_\beta^j\chi_\gamma^k
\chi_\delta^l\, \varepsilon^{\alpha\beta\gamma\delta}
\ee
where $\varepsilon^{\alpha\beta\gamma\delta}$ is the $SU^*(4)$
invariant antisymmetric tensor, and 
we take the Minkowski metric to have `mostly plus'
signature. The supersymmetry transformations are given by
\bea
\delta \phi^X &=& if^X_{ia}\epsilon^{\alpha a} \chi_\alpha^i \\
\delta\chi_\alpha^i&=& f_X^{ia}\partial_{\alpha\beta}\phi^X \epsilon^\beta_a
- \delta\phi^X\omega_X{}^i{}_j\chi_\alpha^j
\eea
where $\epsilon^\alpha_a$ is a constant $Sp_1\times SU^*(4)$ spinor
parameter.

For later use we derive the conditions for a bosonic configuration
of the D=6 supersymmetric sigma model to preserve supersymmetry.
We require the supersymmetry variations of the D=6 fermion
fields to vanish. Written in standard Dirac matrix form, with Lorentz
spinor indices suppressed, this condition becomes
\be\label{normal}
\Gamma^m f_X^{ia}\partial_m\phi^X \epsilon_a =0\, ,
\ee
where $\Gamma^m$ ($m=0,1,\dots,5$) are the D=6 Dirac matrices and
$\epsilon_a$ is an $Sp_n$-Majorana and Weyl spinor; the latter
condition implies that
\be\label{chiral}
\Gamma^{012345}\epsilon = \epsilon\, .
\ee
Multiplying (\ref{normal}) by $f^Y_{ib}$, 
and using the identity (\ref{fiden}), we find
that this is equivalent to  
\be\label{susy6}
\Gamma^m \left[\partial_m \phi^X+ \partial_m\phi^Y i\bfs \cdot
 {\bf I}_Y{}^X\right]\epsilon =0
\ee
where we now suppress the $Sp_1$ indices. Equivalently
\be\label{susypres}
\Gamma^m\epsilon\, \partial_m\phi^Y g_{YX} = i\Gamma^m \bfs \epsilon \cdot
\bfO_{XY}\partial_m\phi^Y\, .
\ee
The dimension of the space of solutions
for constant $\epsilon$ is the number of supersymmetries preserved. 

\subsection{D=4 HK sigma models}

The fermion fields of a D=4 HK sigma model consist of a set of
$2n$ complex 2-component $Sl(2;\bC)$ spinors $\chi_\alpha^i$ ($\alpha=1,2$) and
their complex conjugates $\bar\chi_{\dot\alpha i}$. The general massive sigma
model has a Lagrangian of the form
\bea
{\cal L} &=& {1\over4}g_{XY}\partial^{\alpha\dot
\alpha}\phi^X\partial_{\alpha\dot\alpha}\phi^Y -i\chi_\alpha^i {\cal
D}^{\alpha\dot\alpha} \bar\chi_{\dot\alpha i} 
 + {i\over4}\chi^{\alpha i}\chi_\alpha^j \bar M_{ij}
-{i\over4}\bar\chi^{\dot\alpha i} \bar\chi_{\dot\alpha}^j M_{ij} - V\nn
&&\ - {1\over4}R_{ijkl}(\chi^{\alpha i}\chi_\alpha^j)(\bar\chi^{\dot
\alpha k}\bar\chi_{\dot\alpha}^l) \, ,
\eea
where we assume a `mostly-plus' Minkowski signature and 
use conventions for which
\be
\varepsilon^{\beta\gamma}\varepsilon_{\alpha\gamma} =
\delta_\alpha{}^\beta\, ,\qquad
\partial^{\alpha\dot\beta}\partial_{\alpha\dot\beta} = -2\square\, .
\ee
The complex Yukawa coupling tensor $M_{ij}$, and the potential $V$ are
expressed in terms of a complex vector field $N^X$ appearing 
in the supersymmetry
transformations,
\bea
\delta\phi^X &=& if^X_{ia}\epsilon^{\alpha a}\chi_\alpha^i + c.c. \\
\delta \chi_\alpha^i &=& f_X^{ia}\left[ \partial_{\alpha\dot\alpha}\phi^X
\bar\epsilon^{\dot\alpha}_a + N^X\epsilon_{\alpha a}\right] 
- \delta\phi^X \omega_X{}^i{}_j\chi_\alpha^j \, .
\eea
Specifically, one finds that
\be
M_{ij} = N_{ia;j}{}^a \ ,\qquad V= {1\over2} g_{XY}N^X\bar N^Y\, .
\ee
Supersymmetry implies that $N$ is tri-holomorphic, so that $M_{ij}$
is the only non-vanishing component of $N_{X;Y}$, and $M_{ij}=M_{ji}$. The
{\sl only} other condition required by supersymmetry is that the vector
fields $N$ and $\bar N$ commute.

To obtain these results it is convenient to begin by supposing $M_{ij}$, $N^X$
and $V$ to be dimensionless tensors on the target space and to introduce a mass
parameter $\mu$ (set to unity at the end of the calculation) to take
care of the dimensions. Only variations that vanish when $\mu=0$ need be
considered because the invariance of the massless model is long-established.
Cancellation of the terms linear in $\mu$ determines
$M_{ij}$ and establishes that the complex vector field $N$ must be both
Killing and tri-holomorphic. Cancellation of terms quadratic in $\mu$
determines $V$ {\sl and} requires that $[N,\bar N]=0$. Thus, we have now shown
that the scalar potential of the general `massive' N=2 D=4 sigma model is 
\be
V= {1\over2}g_{XY}\left(N_1^XN_1^Y + N_2^XN_2^Y\right)
\ee
where $g$ is the HK metric and $N_1,N_2$ are two real commuting tri-holomorphic
KVFs. Note that this potential is precisely of the form that one gets by a
non-trivial dimensional reduction from D=6 with
\be\label{dimred4}
\partial_4 \phi = N_1\, ,\qquad \partial_5 \phi = N_2\, .
\ee

\subsection{D=3 HK sigma models}

The D=3 hypermultiplet fermions $\chi_i^A$ transform in the $(2n,2)$
representation of $Sp_n\times Sp_1'$, with $Sp_1\times Sp_1'$ being the
automorphism group of the supersymmetry algebra. The massive N=4 D=3 sigma
model has the Lagrangian (with Lorentz spinor indices suppressed)
\bea
{\cal L} &=&  -{1\over2}g_{XY}\partial^\mu \phi^X \partial_\mu \phi^Y
-{i\over2}\bar\chi^{iA} \gamma^\mu {\cal D}_\mu \chi_{iA} 
+ {i\over4}\bar\chi^{iA}\chi^{jB}M_{ijAB} - V\nn
&&\ -{1\over 12}R_{ijkl}(\bar\chi^{iA}\chi^{Bj})(\bar\chi_A^k\chi_B^l)
\eea
where $\gamma^\mu$ ($\mu=0,1,2)$ are the D=3 Dirac matrices and 
\be
\bar\chi^{iA} = (\chi_{iA})^\dagger\gamma^0
= \Omega^{ij}\varepsilon^{AB}\chi_{jB}^T\gamma^0\, .
\ee
The supersymmetry transformations are
\bea
\delta\phi^X &=& if^X_{ia} \bar\epsilon^{aA}\chi^i_A \\
\delta\chi^i_A &=& f_X^{ia}\left[\gamma^\mu\partial_\mu\phi^X \epsilon_{aA} 
+ N^X{}_A{}^B \epsilon_{aB}\right] - \delta\phi^X\omega_X{}^i{}_j
\chi^j_A\, .
\eea
Supersymmetry fixes the Yukawa tensor to be
\be
M_{ijAB} = (N_{AB})_{ia;j}{}^a
\ee
and the potential to be
\be
V= {1\over4}N^{XAB}N_{XAB}\, .
\ee
We can write
\be
N^X{}_A{}^B = i(\bfs)_A{}^B \cdot {\bf k}^X
\ee
where ${\bf k}$ is a triplet of vector fields. Supersymmetry implies that they
are Killing, tri-holomorphic, and mutually commuting.  Thus, the general
sigma model potential compatible with N=4 D=3 supersymmetry is
\be
V= {1\over2}g_{XY}{\bf k}^X\cdot {\bf k}^Y 
\ee
where $g$ is the HK metric and ${\bf k}$ is a triplet of mutually commuting
tri-holomorphic KVFs. This is again precisely of the form that one would get by
non-trivial dimensional reduction from D=6. 

\section{Static supersymmetric solitons}

We now  focus on the toric HK manifolds. Their essential features were
summarized in the introduction. We take $n\ge2$ and consider a scalar potential
of the form (\ref{pottwo}). As we have just seen, this potential is compatible
with both D=4 N=2 and D=3 N=4 supersymmetry. For generic mass-squared 
matrix $\mu^2$ the
potential vanishes only when the matrix $U^{IJ}$ vanishes. This
yields a set of discrete supersymmetric vacua. We are interested in the kinks
that interpolate between these vacua, which we expect to preserve 1/2
supersymmetry. In D=3,4, kinks are domain walls and, as we shall see, we can
have intersecting walls that preserve 1/4 supersymmetry. 

\subsection{Energy bounds}

Consider the case of time-independent fields. The D=3 energy density, 
which is the same as the D=4 energy density for fields that
are translational-invariant in the 3-direction, is then  
\bea
{\cal E} &=& {1\over2}U_{IJ} \nabla_1{\bf X}^I\cdot \nabla_1{\bf X}^J
+ {1\over2}U_{IJ} \nabla_2{\bf X}^I\cdot \nabla_2{\bf X}^J
+ {1\over2}\mu_{IJ}U^{IJ} \nn
&+&{1\over2}U^{IJ}{\cal D}_1\psi_I {\cal D}_1\psi_J
+{1\over2}U^{IJ}{\cal D}_2\psi_I {\cal D}_2\psi_J
\eea
where
\be
{\cal D}_\mu\psi_I=\nabla_\mu\psi_I+\nabla_\mu{\bf X}^K\cdot{\bfo}_{KI}
\ee
We can rewrite this as
\bea
{\cal E} &=& {1\over2}U_{IJ}\left( \nabla_1{\bf X}^I - U^{IK}{\bf n}_K\right)
\left(\nabla_1{\bf X}^J - U^{JL}{\bf n}_L\right) + \nabla_1 {\bf X}^I\cdot
{\bf n}_I \nn
\ &+& {1\over2}U_{IJ}\left( \nabla_2{\bf X}^I - U^{IK}{\bf p}_K\right)
\left(\nabla_2{\bf X}^J - U^{JL}{\bf p}_L\right) + \nabla_1 {\bf X}^I\cdot
{\bf p}_I \nn
\ &+&  {1\over2}U^{IJ}{\cal D}_1\psi_I {\cal D}_1\psi_J
+{1\over2}U^{IJ}{\cal D}_2\psi_I {\cal D}_2\psi_J
\eea
where ${\bf n}_I$ and ${\bf p}_I$ are constants such that
\be
{\bf n}_I \cdot {\bf n}_J + {\bf p}_I \cdot {\bf p}_J = \mu^2_{IJ}\, .
\ee
Note that the integrals $\int dx^1\nabla_1{\bf X}^I$ and $\int dx^2\nabla_2 {\bf
X}^I$  are topological charges carried by domain wall orthogonal to the
1-direction and the 2-direction, respectively. From the above expression for 
the energy density we deduce the bound
\be\label{rain}
{\cal E} \ge \nabla_1 {\bf X}^I\cdot {\bf n}_I + \nabla_2 {\bf X}^I\cdot {\bf
p}_I\, .
\ee
This is saturated by solutions of
\be\label{toricbog}
\nabla_1 {\bf X}^I = U^{IJ}{\bf n}_J\, ,\qquad
\nabla_2 {\bf X}^I = U^{IJ}{\bf p}_J\, ,
\ee
together with 
\bea\label{BOGTWO}
\nabla_1 \psi_I = - U^{JK} \bfo_{JI}\cdot {\bf n}_K \, ,\qquad
\nabla_2 \psi_I = - U^{JK}\bfo_{JI}\cdot {\bf p}_K\, ,
\eea
which can be solved for $\psi_I$ given a
solution of (\ref{toricbog}). Consistency of these equations requires that
\be\label{consist}
{\bf n}_{(I} \times {\bf p}_{J)} =0\, .
\ee

\subsection{Supersymmetry}

We wish to determine the fraction of supersymmetry preserved by solutions of
(\ref{toricbog}) and (\ref{BOGTWO}). This could be done directly from the D=3,4
supersymmetry transformation laws, but it can also be done from the
supersymmetry  transformations of the D=6 massless sigma
model, by making use of the fact that the bosonic part of
the massive D=4 model is obtained by the reduction ansatz
\be\label{massans}
\partial_4 \phi^X = (\lambda_Ik^I)^X\, ,\qquad \partial_5\phi^X =
(\rho_Ik^I)^X\, .
\ee
This is just (\ref{dimred4}) with $N_1= \lambda_Ik^I$ and $N_2=\rho_Ik^I$. 
A further trivial dimensional reduction leads to the D=3 model
with the same scalar potential.

Substituting (\ref{massans}) into the D=6 supersymmetry preservation condition
(\ref{susypres}), and using the explicit form of the metric and Kahler 2-forms
for toric HK manifolds, we obtain the two conditions
\be\label{first}
\left[\Gamma^m\bfs\cdot \partial_m {\bf X}^I + iU^{IJ}\Gamma^m
{\cal D}_m\psi_J\right]\epsilon =0
\ee
and 
\be\label{second}
\left[\Gamma^m\partial_m{\bf X}^I -i\Gamma^m\bfs \times
\partial_m{\bf X}^I +iU^{IJ}\Gamma^m\bfs {\cal
D}_m\psi_I\right]\epsilon =0\, .
\ee
The second of these conditions is actually implied by the first, so that
all we need consider is (\ref{first}). Using (\ref{massans}) 
this becomes
\be\label{susypres2}
\Gamma^\mu \left[\bfs \cdot \partial_\mu {\bf X}^I +iU^{IJ} {\cal D}_\mu
\psi_J \right]\epsilon =  -iU^{IJ}[\lambda_J\Gamma^4 + \rho_J\Gamma^5] \epsilon
\ee
where, $\mu=0,1,2,3$. This is the condition for partial preservation of
supersymmetry for the massive sigma models under consideration.
When applied to time-independent, and $x^3$-independent, configurations
satisfying the BSTPS equations (\ref{toricbog}) and (\ref{BOGTWO}), 
the condition (\ref{susypres2}) becomes
\be\label{gencon}
\left(\Gamma^1\bfs \cdot {\bf n}_J + \Gamma^2\bfs\cdot {\bf p}_J\right)\epsilon
= -i\left(\lambda_J\Gamma^4 + \rho_J \Gamma^5\right)\epsilon
\ee
Using (\ref{consist}), we deduce the consistency condition
\be\label{consist2}
\bfs \cdot\left({\bf n}_1\times {\bf n}_2 + {\bf p}_1\times {\bf
p}_2\right)\epsilon -i\Gamma^{12}\left({\bf n}_1\cdot {\bf p}_2 - {\bf n}_2
\cdot {\bf p}_1\right) \epsilon -i\Gamma^{45} \left(\lambda_1\rho_2
-\lambda_2\rho_1\right)\epsilon =0\, .
\ee
There are various ways to ensure that both this condition and
(\ref{consist}) are satisfied. One is to set
\be\label{choice}
{\bf n}_I = {\bf n}\lambda_I \, ,\qquad {\bf p}_I = {\bf n}\rho_I
\ee
for unit 3-vector ${\bf n}$. In this case (\ref{consist}) is
identically satisfied and (\ref{consist2}) implies {\sl either} that
$\lambda_I$ and $\rho_I$ are proportional, {\sl or} that
\be\label{4gamma}
\Gamma^{1245}\epsilon =\epsilon\, .
\ee

We shall make the choice (\ref{choice}) in what follows, in 
which case the equations
(\ref{toricbog}) and (\ref{BOGTWO}) reduce to
\be\label{BPS}
\nabla_1 {\bf X}^I = {\bf n}\, U^{IJ}\lambda_J\, ,\qquad
\nabla_2 {\bf X}^I = {\bf n}\, U^{IJ}\rho _J\, .
\ee
and 
\be\label{BPSTWO}
\nabla_1\psi_I = - U^{JK}\lambda_K \bfo_{JI}\cdot {\bf n}\,, \qquad
\nabla_2\psi_I = - U^{JK}\rho_K \bfo_{JI}\cdot {\bf n}\, .
\ee
It is straightforward to show that generic solutions of these equations
preserve 1/4 supersymmetry. We simply note that the choice (\ref{choice}) 
reduces (\ref{gencon}) to
\be\label{gencontwo}
\Gamma^4\lambda_I\left[1+ i\Gamma^{14}(\bfs\cdot{\bf n})\right] \epsilon + 
\Gamma^5\rho_I \left[1+ i\Gamma^{25}(\bfs\cdot{\bf n})\right] \epsilon =0\, .
\ee
If $\lambda_I$ and $\rho_I$ are not proportional we must impose
both (\ref{4gamma}) and
\be\label{conone}
-i\Gamma^{14}(\bfs\cdot{\bf n})\epsilon = \epsilon\, . 
\ee
These two constraints are compatible and imply preservation of 1/4
supersymmetry. If $\lambda_I$ and $\rho_I$ are proportional then the constraint
(\ref{4gamma}) is not needed and (\ref{gencontwo}) implies
preservation of 1/2 supersymmetry. An example is the kink domain wall
with $\rho=0$ but $\lambda\ne 0$, in which case the only constraint on
$\epsilon$ is (\ref{conone}). 

\subsection{Intersecting domain walls}

We now aim to show that the equations (\ref{BPS}) and (\ref{BPSTWO}) admit
solutions that can be interpreted as intersecting domain walls. 
When considering the toric HK 8-metrics it is convenient to set
\be
{\bf X}^I = ({\bf X},{\bf Y}) \, . 
\ee
We shall focus here on the special case in which the $2\times 2$ matrix
function $U$ takes the form
\be
U_{IJ}= \pmatrix{U({\bf X})&c\cr c&U ({\bf Y})}
\ee
for constant $c$ (such that $\det U\ne0$). For this choice
\be
A_1= d{\bf X}\cdot {\bf A}({\bf X}) \, 
\qquad A_2= d{\bf Y}\cdot {\bf A}({\bf Y})
\ee
with $\bfn \times {\bf A} = \bfn U$. It follows that $U$ is a harmonic
function. We shall choose 
\be
U({\bf Z})=  a + {1\over2}\left[ {1\over |{\bf Z}-{\bf n}|} + 
{1\over |{\bf Z}+{\bf n}|}\right]
\ee
for ${\bf Z}=({\bf X},{\bf Y})$ and unit 3-vector ${\bf n}$. 
The factor of $1/2$ ensures that
the metric is complete if the angles $\psi_I$ have period $2\pi$. When
$c=0$ the matrix $U$ is diagonal and the 8-metric is the direct sum of two
identical two-centre HK 4-metrics, but the 8-metric is irreducible 
when $c\ne0$. 

If we identify the unit 3-vector ${\bf n}$ with the unit 3-vector appearing in
(\ref{choice}) and choose ${\bf n}\cdot {\bf A}=0$, as is
possible \cite{paptownsend}, then (\ref{BPSTWO}) becomes
\be
\nabla_x \psi_I = \nabla_y \psi_I =0
\ee
where we have renamed the two space coordinates $(x^1,x^2)$ as
$(x,y)$.
These equations imply that the angles $\psi_I$ are constants, which
play no further role in the solution. After setting
\be
{\bf X} = {\bf n}X\, ,\qquad {\bf Y}= {\bf n}Y \, ,
\ee
the remaining BPS equations (\ref{BPS}) can be written as
\be\label{redbog}
\pmatrix{U(X) \nabla_x X + c\nabla_x Y &
U(Y) \nabla_x Y + c\nabla_x X \cr
U(X) \nabla_y X + c\nabla_y Y &
U(Y) \nabla_y Y + c\nabla_y X }
= \pmatrix{\lambda_1 & \lambda_2\cr \rho_1 & \rho_2}\, .
\ee
These equations can be solved by setting
\be
X= \tanh u\, ,\qquad Y= \tanh v\, ,
\ee
and the solution is then given implicitly by
\bea\label{abeqs}
u + a\tanh u + c\tanh v &=& \lambda_1 x + \rho_1 y\nn
v + a\tanh v + c\tanh u &=& \lambda_2 x + \rho_2 y
\eea
Equivalently, $X(x,y)$ and $Y(x,y)$ are given implicitly by
\bea\label{advert}
X= \tanh \left[ \lambda_1 x + \rho_1 y - aX -cY\right]\nn
Y= \tanh \left[ \lambda_2 x + \rho_2 y - aY -cX\right]
\eea
This is the advertized exact solution. If $\lambda_I$ and $\rho_I$
are proportional then $X$ and $Y$ are independent of one linear combination of
$x$ and $y$, so the solution is a domain wall (shown above to preserve 1/2
supersymmetry). Otherwise we have a 1/4 supersymmetric solution representing
the intersection of two domain walls.

For simplicity we shall now set $a=0$ and choose
\be
\lambda_I=(1,0) \, ,\qquad \rho_I = (0,1)\, ,
\ee
corresponding to the potential
\be
V= {1\over2}\left(|k^1|^2  + |k^2|^2\right)\, .
\ee
In this case (\ref{advert}) reduces to 
\bea\label{advert2}
X= \tanh (x -cY)\, ,\qquad  Y= \tanh (y -cX)
\eea
For $c=0$ this reduces further to $X= \tanh x$ and $Y=\tanh y$, i.e., two
orthogonally intersecting planar domain walls. The intersection is trivial, as
expected, because for $c=0$ we have the sum of two massive sigma models,
each with an identical Eguchi-Hansen target space 4-metric. For $c\ne0$ 
the two domain walls are asymptotically planar but are deformed 
near the intersection, as we shall now show. 

The energy density for the solution (\ref{advert2}) is
\be
{\cal E}(x,y) = {\cosh^2 u + \cosh^2 v\over \cosh^2 u\cosh^2 v -c^2}\, .
\ee
This has a maximum at $x=y=0$, which we may take to be the `point' of
intersection of the domain walls. This maximum occurs at $u=v=0$, and
\be
{\cal E}_{max} = {2\over 1-c^2}\, .
\ee
As the non-vanishing of $\det U$ requires $|c|<1$ (when $a=0$, as we are
now assuming) this maximum is finite and positive. The angle $\theta$ between
the domain walls at the point of intersection is given by
\be
\cos\theta = {\bfn X\cdot \bfn Y \over |\bfn X||\bfn Y|}\big|_{u=v=0}
= {\bfn u\cdot \bfn v \over |\bfn u||\bfn v|}\big|_{u=v=0}\, ,
\ee
which yields
\be\label{angle}
\cos\theta =  -2\ {c\over 1+c^2}\, .
\ee
Note that $\theta$ is real, as it should be, because $|c|<1$. 

Although the domain walls are not orthogonal near the intersection point,
they are asymptotically orthogonal.
One can get an idea of the shape of the solution by rewriting  (\ref{advert2})
as
\bea
u + c\tanh \left(y- c\tanh u\right) &=& x\nn
v + c\tanh\left(x-c\tanh v\right) &=& y
\eea
The curves in the $(x,y)$ plane corresponding to $u=0$ and $v=0$ (which are the
curves on which ${\cal E}$ is a maximum for fixed $v$ and $u$,
respectively) are
\be
x= c\tanh y\, ,\qquad y= c\tanh x\, ,
\ee
respectively. 
For $c=0$ these are just the $(x,y)$ coordinate axes, as expected, but for
$c\ne0$ they each shift by $2c$ in passing through the origin, just so as to
intersect at the angle $\theta$ given by (\ref{angle}).

\section{Stationary supersymmetric solitons}

We shall say that a non-static field configuration is stationary if
its energy density is time-independent. The Q-kinks of \cite{AT2}
provide simple examples of stationary but non-static domain walls in
models with 4-dimensional HK target spaces. These can be generalized
to 1/2 supersymmetric domain wall solutions of models with
$4n$-dimensional toric HK target spaces, as we shall see below, but
our principal aim in this section is to exploit the new
features provided by higher-dimensional target spaces to find
essentially different stationary generalizations of the intersecting
domain wall solutions discussed in the previous section.   

In the case of intersecting domain walls of the N=1 D=4 Wess-Zumino
model it was found to be possible to add a wave along the intersection,
preserving 1/4 supersymmetry \cite{GT}. One can ask whether the same is true of
the N=2 D=4 sigma model domain walls. Given such a configuration, its
reduction to D=3 would be expected to yield a charged intersecting
domain wall. We shall now show that such D=3 configurations,
preserving 1/4 supersymmetry indeed exist, provided the target space
admits at least {\sl three} linearly independent tri-holomorphic KVFs.
For this we need at least a 12-dimensional toric HK target space. Our starting
point will be a massive D=3 N=4 supersymmetric sigma model with potential
\be
V= {1\over2} \mu^2_{IJ}U^{IJ}
\ee
where
\be
\mu^2_{IJ}= (\lambda_I\lambda_J + \rho_I\rho_J + q_Iq_J)\, .
\ee
This potential can be obtained by reduction from D=6 with
\be
\partial_4 \psi_I = \lambda_I\qquad
\partial_5 \psi_I = \rho_I \qquad
\partial_3 \psi_I = q_I
\ee
where $\lambda,\rho,q$ are three 3-vectors. 

\subsection{Charged intersecting domain walls}

The energy density of the above model can be written in the form 
\bea
{\cal E} &=&  {1\over2}U_{IJ}\left( \nabla_1{\bf X}^I -
U^{IK}\lambda_K
{\bf n}\right)
\left(\nabla_1{\bf X}^J - U^{JL}\lambda_L{\bf n}\right) + 
\nabla_1 X^I\lambda_I \nn
\ &+& {1\over2}U_{IJ}\left( \nabla_2{\bf X}^I - U^{IK}\rho_K{\bf n}\right)
\left(\nabla_2{\bf X}^J - U^{JL}\rho_L{\bf n}\right) + 
\nabla_2 X^I\rho_I \nn
&+&{1\over2}U^{IJ}{\cal D}_1\psi_I {\cal D}_1\psi_J
+{1\over2}U^{IJ}{\cal D}_2\psi_I {\cal D}_2\psi_J
+ {1\over 2}U_{IJ}{\dot{\bf X}}^I\cdot{\dot{\bf X}}^J\nn
&+&{1\over2}U^{IJ}({\cal D}_t\psi_I \mp q_I)( {\cal D}_t\psi_J\mp q_J)
\pm U^{IJ}{\cal D}_t\psi_I q_J
\eea
where ${\bf n}$ is a constant unit 3-vector and 
$X={\bf X}\cdot {\bf n}$. Noting that $\int d^2x \bfn X^I$ are 
topological kink
charges and $\int d^2x\, U^{IJ}{\cal D}_t \psi_J$ are 
the Noether charges
corresponding to the isometries generated by the Killing vector fields $k^I$, 
we deduce the bound
\be\label{rain2}
{\cal E} \ge \nabla_1 X^I\lambda_I + \nabla_2 X^I\rho_I + 
U^{IJ}{\cal D}_t\psi_I q_J\, ,
\ee
which is saturated when 
\be\label{sprain}
\qquad \dot X^I=0\, \qquad \dot\psi_I = \pm q_I\, ,
\ee
together with 
\be
{\cal D}_1\psi^I= {\cal D}_2\psi^I=0\, ,
\ee
and 
\be\label{sprainth}
\nabla_1 {\bf X}^I =  U^{IJ}\lambda_I{\bf n}\, \qquad 
\nabla_2 {\bf X}^I = U^{IJ}\rho_J{\bf n}\, .
\ee

The simplest way to solve these equations is to suppose the target
space to be a direct product of the 8-dimensional HK manifold
chosen previously with a 4-dimensional HK manifold for which 
$k^3= \partial/\partial\psi_3$ is the tri-holomorphic KVF. If one then
chooses $q_1=q_2=0$ (which one can do without loss of generality) and 
$\lambda_3=\rho_3 =0$ then the solution is
identical to the previous one but with $\psi_3 = q_3 t$. It seems
likely that this is a limiting solution of a more general one for
irreducible 12-dimensional HK target spaces, but we shall not pursue
this here. Assuming that we have a solution, we now turn to the
determination of the fraction of supersymmetry it preserves. 

The supersymmetry preservation condition for the massive D=3 models under
consideration is
\be\label{susypres3}
\left[\bfs \cdot {\bf n}\, \Gamma^\mu \partial_\mu X^I\right]\epsilon = 
-iU^{IJ}[\lambda_J\Gamma^4 + \rho_J\Gamma^5 + 
q_J \Gamma^3 + \Gamma^\mu D_\mu \psi_J] \epsilon\, .
\ee
where, now, $\mu=0,1,2$. For solutions of the first-order equations 
(\ref{sprain})-(\ref{sprainth}) this becomes
\be\label{epcon}
\big[\left(\Gamma^1 \bfs \cdot {\bf n} + i\Gamma^4\right) \lambda_I
+ \left(\Gamma^2 \bfs \cdot {\bf n} + i\Gamma^5\right)\rho_I 
+i \left(\Gamma^3 \pm \Gamma^0\right) q_I\big]\epsilon =0\, .
\ee
If $q$ vanishes but $\lambda$ and $\rho$ are not proportional
then supersymmetry imposes the conditions
\be\label{quart}
i\Gamma^{51}\bfs\cdot {\bf n}\, \epsilon = \epsilon\, , \qquad
i\Gamma^{42}\bfs\cdot {\bf n}\, \epsilon = \epsilon\, ,
\ee
which preserve 1/4 supersymmetry. Because of the chirality condition
(\ref{chiral}), these conditions imply that  $\Gamma^{03}\epsilon=\epsilon$, so
that 1/4 supersymmetry is maintained when $q_I$ is non-zero and
$\dot \psi_I = q_I$ but all supersymmeties are broken if $\dot\psi_I =-q_I$. 

We conclude that 1/4 supersymmetric charged intersecting domain walls are
possible in this D=3 model. They are stationary but not static. Intersecting
domain walls fitting this description can also be found rather more easily by
replacing one or both of the kink domain walls by a Q-kink domain
wall, but the solution just found is of a different type that is
possible only when the toric HK manifold has dimension $4n$ with
$n\ge3$. It has the feature that the charge can be interpreted in D=4 as a wave
with the speed of light along the intersection. This follows directly from the
fact that the D=4 fields $\psi_I$ satisfy 
\be
(\partial_t - \partial_3)\psi_I =0\, , 
\ee
since the D=3 model is obtained from D=4 by the ansatz $\partial_3
\psi_I = q_I$. Note that a wave in the opposite direction breaks
all the supersymmetry, so the intersection is chiral. It can viewed as
a (2,0) supersymmetric string. 

\subsection{Special Cases}

The mass-squared matrix $\mu^2$ in the models just considered has
maximal rank three and the generic examples will occur when the rank
is precisely three. We now wish to consider some special cases of
stationary supersymmetric solitons that arise when $\mu^2$ has rank
less than three. Consider the rank two case, which we can arrange by
choosing $\rho=0$. We again make the ansatz ${\bf X}^I=
X^I{\bf n}$, and we may now set $\nabla_y X^I=0$. We are then left with the
equations
\be\label{eqsstat}
\nabla_x X^I = U^{IJ}\lambda_J\, ,\qquad
\dot\psi_I = \pm q_I
\ee
to be solved for $X^I(x)$ and (trivially) for $\psi_I(t)$. 
The supersymmetry preservation condition is now
\be\label{finalpres}
\lambda_I\left[i\Gamma^4+\Gamma^{1}(\bfs\cdot{\bf n})\right] \epsilon +
iq_I\left[\Gamma^0 \pm\Gamma^{3}\right] \epsilon =0\, ,
\ee
which implies preservation of 1/4 supersymmetry when $\mu^2$
has rank two. Solutions of (\ref{eqsstat}) thus yield new 1/4
supersymmetric charged domain walls, and since $\rho=0$ these are also
solutions of the D=4 massive sigma model.  

If instead $\mu^2$ has rank one then (\ref{finalpres}) implies
preservation of 1/2 supersymmetry. To see this we set
\be
q_I = v\nu_I\, ,\qquad
\lambda_I = \sqrt{1-v^2}\, \nu_I\, ,
\ee
where $\nu_I$ are constants (not all zero) and $v$ is another constant
with $|v|<1$. In this case (\ref{finalpres}) is equivalent to 
\be
\Gamma\epsilon = \epsilon
\ee
with
\be
\Gamma = i\Gamma^{41} \bfs \cdot {\bf n} + {v\over
\sqrt{1-v^2}}\left(\Gamma^{04} \pm \Gamma^{34}\right)
\ee
Since $\Gamma$ is traceless and satisfies $\Gamma^2=1$, this condition
implies preservation of 1/2 supersymmetry. The equations that yield
these 1/2 supersymmetric stationary domain walls are
\be
\nabla_x X^I = \sqrt{1-v^2}\, U^{IJ} \nu_J\, ,\qquad \dot\psi_I = \pm
v \nu_I\, .
\ee
Solutions of these equations generalize the Q-kinks of \cite{AT2} to
higher-dimensional HK target spaces. 

\section{Discussion}

We have shown that massive hyper-Kahler D=4 supersymmetric sigma models can 
admit non-trivial static intersecting domain wall solutions that preserve 1/4 of
the supersymmetry, provided that the target space is at least 8-dimensional and
admits at least two linearly independent commuting tri-holomorphic Killing
vector fields. Such models also admit new stationary charged domain wall
solutions that also preserve 1/4 supersymmetry. Given three linearly
independent tri-holomorphic Killing vector fields, which requires a target
space of at least 12 dimensions, there exist 1/4 supersymmetric stationary
intersecting domain wall solutions in D=4 with a wave along the string
intersection. Reduction on the string direction yields a charged intersection
domain wall solution of a massive D=3 sigma-model, again preserving 1/4
supersymmetry. We have found exact solutions for a class of sigma
models with particular toric hyper-K\"ahler target spaces. It is unlikely that
{\sl exact} solutions can be found in general, but we 
expect intersecting domain wall solutions to {\sl exist}  for many
other toric HK 8-manifolds too, e.g. the Calabi manifolds
that we discuss below.

Whereas 1/4 supersymmetry is the minimal possible fraction of N=1 D=4
supersymmetry it is twice the minimal possible fraction of N=2 D=4 (or N=4 D=3)
supersymmetry. One wonders whether there are other intersecting brane
configurations that preserve only 1/8 supersymmetry. It is not difficult to
find such solutions in models with factorizable target spaces because one can
just superpose solutions within each factor. It would be interesting to
know if these can be generalised to non-trivial examples. 

Although we have focussed on solutions of D=3,4 models, we note that
the supersymmetric charged domain wall solutions discussed in
section 4.2 give rise to supersymmetric kink solutions in D=2.
We now conclude with a brief discussion of a 
possible applications of these solutions to IIB superstring theory. 
Consider a single D1-brane in the presence 
of k parallel D5-branes. The low-energy effective dynamics of the D-string is
given by a  two dimensional quantum field theory which, for coincident
D5-branes, has both a Coulomb and a Higgs branch. Switching on a constant NS
B-field parallel to the D5-branes corresponds to switching on Fayet-Illiopoulos
terms in the  two-dimensional quantum field theory on the D-string, and this
lifts the Coulomb branch. In this case the Higgs branch is the charge-1
non-commutative instanton moduli space (after ignoring the centre of mass). In
other words, the low-energy dynamics of the Higgs branch is given by a D=2
supersymmetric HK sigma model with a 
$4(k-1)$-dimensional Calabi space as target
\cite{leeyi}. This space has $k-1$ mutually commuting tri-holomorphic Killing
vector fields. Separating the D5-branes along an axis 
leads to the addition of
a supersymmetric preserving potential 
constructed  from a single tri-holomorphic
Killing vector field. The 1/2 supersymmetric kink and Q-kink solutions were
interpreted in \cite{TL}, following \cite{bergtown}, as (1,Q) strings
interpolating from one D5-brane to another. For k=3, when the D5-branes are
not co-linear, the potential is constructed from two tri-holomorphic
Killing vector fields. It seems likely that the 1/4 supersymmetric 
charged kink solutions found here can be interpreted as a string 
that interpolates from one D5-brane to two D5-branes via a string junction.

\vspace{.5truecm}

\noindent {\bf Acknowledgements} 
DT would like to thank the Center for Theoretical Physics, MIT, and 
Sunil Mukhi and the Department of Theoretical Physics, TIFR for 
their hospitality. JPG thanks the EPSRC for partial support. DT is 
supported by an EPSRC fellowship. All authors are supported in part by 
PPARC through their SPG $\#$613.

\end{document}